# Optimizing Energy Efficiency of Wearable Sensors Using Fog-assisted Control

Delaram Amiri[1]    Arman Anzanpour[2]    Iman Azimi[2]
Amir M. Rahmani[1]    Pasi Liljeberg[2]    Nikil Dutt[1]
Marco Levorato[1]

[1]University of California Irvine, USA
[2]University of Turku, Finland

**Abstract**

Recent advances in the Internet of Things (IoT) technologies have enabled the use of wearables for remote patient monitoring. Wearable sensors capture the patient's vital signs, and provide alerts or diagnosis based on the collected data. Unfortunately, wearables typically have limited energy and computational capacity, making their use challenging for healthcare applications where monitoring must continue uninterrupted long time, without the need to charge or change the battery. Fog computing can alleviate this problem by offloading computationally intensive tasks from the sensor layer to higher layers, thereby not only meeting the sensors' limited computational capacity but also enabling the use of local closed-loop energy optimization algorithms to increase the battery life. Furthermore, the patient's contextual information – including health and activity status – can be exploited to guide energy optimization algorithms more effectively. By incorporating the patient's contextual information, a desired quality of experience can be achieved by creating a dynamic balance between energy-efficiency and measurement accuracy. We present a run-time distributed control-based solution to find the most energy-efficient system state for a given context while keeping the accuracy of decision making process over a certain threshold. Our optimization algorithm resides in the Fog layer to avoid imposing computational overheads to the sensor layer. Our solution can be extended to reduce the probability of errors in the data collection process to ensure the accuracy of the results. The implementation of our fog-assisted control solution on a remote monitoring system shows a significant improvement in energy-efficiency with a bounded loss in accuracy.

## Contents

# 1 Introduction

Continuous clinical-level monitoring of patients conditions is imperative in an ample range of medical applications. For instance, monitoring post-operative patients to detect early signs of health deterioration can significantly improve care outcomes. Current technologies can only provide clinical-level monitoring in hospital settings, where the patient is in a controlled environment and traditional sensors can be used. However, once discharged, patients are left in a vulnerable position. Achieving clinical-level monitoring in everyday settings would have a tremendous impact on patients health, but is a technological challenge that has not been solved yet.

Internet of Things (IoT) technologies have been recently widely used to build systems capable of continuously monitoring subjects, acquiring a variety of biosignals [24, 41, 25]. The healthcare IoT paradigm proves a way to ubiquitous and personalized monitoring of individual's conditions in everyday settings. However, these technologies have several limitations that make continuous high-quality sensing difficult. In fact, these sensors have limitations in terms of storage, computation load and energy supply. Furthermore, different from hospital environments, in everyday settings the activities the monitored subject engages may cause a degradation of the quality of the signals due to the movement between the skin and the sensor. Some wearable sensors, such as, Photoplethysmogram (PPG), Electrocardiogram (ECG), and Electromyography (EMG) are particularly influenced by this effect and may suffer a significant loss of accuracy.

Interestingly, different activities may cause a different degree of degradation. For instance, motion artifacts affecting the measured signal when the subject is "Running" are much larger compared to those generated when the person is "Sitting" or "Sleeping". Achieving the same Signal-to-Noise Ratio (SNR) in all the activities and, thus, the same detection quality, requires the tuning of the sensing power to that of the noise. In other words, higher sensing energy could be used when the patient is "Running", and a lower energy could be used when the patient is "Sitting". Nonetheless, wearable sensors are manufactured in industry for worst case scenario in terms of noise level, corresponding to a large sensing power, which results in a high energy consumption and a short battery lifetime. The layered and pervasive IoT infrastructure can be used to support solutions enabling the adaptation of sensing parameters to the joint person-technology system state. The ability of the system to track its own state and determining optimal parameters can be connected with the general concept of "self-awareness". Herein, we extend this notion across the layers of the IoT infrastructure, and focus patient's activity as a main driver of adaptation due to the specific application domain.

In the computing literature, the general notion of context captures the state of the system, including any descriptor of the user's state. User's activity is considered as a subclass of context information, whose estimation and track-ing requires the acquisition of signals from sensors and the implementation of inference algorithms. Intuitively, imposing a further burden on the sensor to have them connect with each other and implement algorithms is not a suitable option. Therefore, the needed "semantic" support necessarily comes from the higher layers of the IoT infrastructure, which has sufficient resources to perform signal fusion and processing. In particular, fog and edge computing resources [36, 37], which are connected to the mobile devices through 1-hop low latency wireless links, are particularly indicated to host compute-intense tasks informing system-level control. Implementing optimization algorithms controlling the sensor layer on the fog layer can avoid imposing overheads to the sensor layer while being able to rapidly respond to changes thanks to the local control. I In this chapter, we propose approaches using the real-time connection between the sensors and the fog layer to build a context-aware and self-aware control loop determining the sensing power used by wearable sensors.

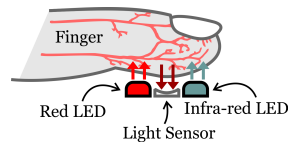

Fig. 1: PPG sensors consisting of two light sources and one light sensor

The rest of the chapter is organized as follows. Section 2 provides a background in remote patient monitoring. Section 3 discusses researches done in energy efficient sensor control. Section 4 discusses the design challenges in healthcare IoT. Proposed energy-efficient fog computing is defined in section 6. Section 7 concludes the chapter.

## 2 Background

Remote health monitoring is a promising approach to extend reactive and proactive healthcare solutions for populations at-risk beyond traditional clinical settings. Such a service allows continuous monitoring of patients in their daily routines, enabling early intervention services in case of health deterioration [8, 41]. Moreover, it has the potential to alleviate medical costs and hospital visits for patients, improving their quality of life as well as independent living. Internet-of-Things (IoT) as an advance network of objects can be advantageously applied in such applications.

IoT-based systems leverage a variety of sensors, communication infrastructures and computing resources to deliver monitoring solutions [21, 12, 24]. In the context of remote health monitoring in every day settings, these systems demand continuous data acquisition with high-level quality attributes, where various vital signs should be collected seamlessly while end-users are involved in daily routines.

In this context, Photoplethysmography (PPG) is a promising non-invasive mechanism to capture various vital signs for users in every day settings. PPG is a non-obtrusive optical method employed to measure blood volume variations in the microvascular bed of tissues [2]. The blood variations are associated with cardiac and respiratory activities. They also reflects an estimate of arterial oxygen saturation. Accordingly, the PPG method is widely used in several wearable and IoT-based systems, through which different vital signs such as heart rate, respiration rate and oxygen saturation (i.e., SpO$_2$) are extracted. The PPG method often contains two light sources along with one light sensor, placed on a body organ such as fingertip (see Figure 1). The light sources emit red and infrared lights to the tissue, while the light sensors absorb the light reflection from the tissue, capturing the PPG waveforms which correspond to the amount of oxygenated Hemoglobin molecules in the veins.

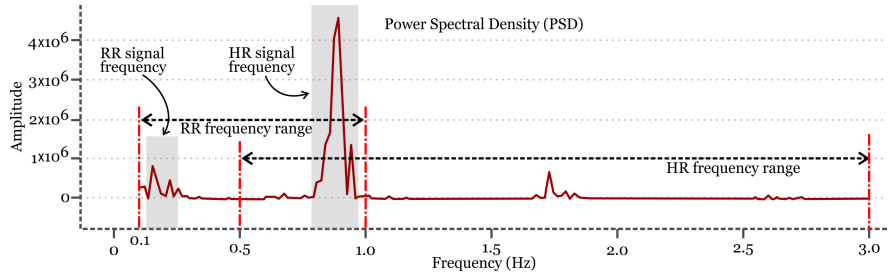

Fig. 2: Power Spectral Density (PSD) of one-minute PPG signal

Different techniques have been proposed to obtain heart rate and respiration rate from the PPG waveform (i.e., red or infrared waveform) [14, 33]. Feature-based techniques are designed to obtain the vital signs by extracting certain features (e.g., maximum intensity of the pulse and baseline variations) from the signal [26]. These techniques, nevertheless, are susceptible to the presence of motion artifacts and surrounding noises which distort the features.

Alternatively, filter-based techniques extract the vital signs, leveraging band-pass filters [20]. In this context, two band pass filters are designed according to the respiration and heartbeat frequency ranges. Initially, the cutoff frequencies are set to 0.1-1 Hz (6-60 breath rate/minute) and 0.5-180 Hz (30-180 beats/minute) for the respiration rate and heart rate extraction, respectively. The boundaries are, then, narrowed down to the vital signs' frequencies to mitigate the noise. In this regard, the cutoff frequencies are dynamically selected exploiting the peak values in the power spectral density (PSD) of the signal [30] (see Figure 2). Leveraging the band-pass filters, the respiratory and heartbeat signals are extracted. Then, a peak detection algorithm is performed to obtain local maximum points in the derivative of the bio-signals. The time interval between two consecutive peaks indicate the heart rate and respiration rate values.

Despite the heart rate and respiration rate, the SpO$_2$ is derived from both infrared and red waveforms. As shown in Figure 3, four features (i.e., $AC_{IR}$, $AC_{RED}$, $DC_{IR}$ and $DC_{RED}$) are first extracted. Then, the SpO$_2$ is determined using the following equations:

$$R = \frac{AC_{RED}.DC_{IR}}{AC_{IR}.DC_{RED}}, \qquad (1)$$

$$SpO_2 = \alpha R^2 + \beta R + \gamma \qquad (2)$$

where $\alpha$, $\beta$ and $\gamma$ are constants obtained from the sensor's specification [32].

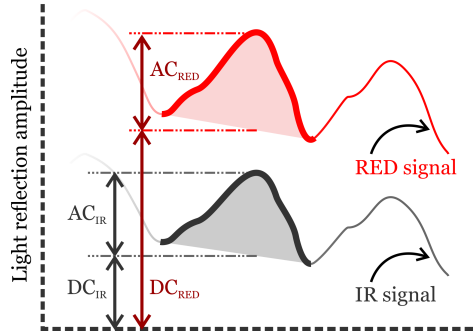

Fig. 3: PPG waveforms and the four features extracted for SpO$_2$ calculation

## 3 Related Topics

In healthcare IoT applications, sensors are frequently energy bottleneck of the system. Algorithms to reduce energy consumption of sensors have been widely studied. Different approaches have been proposed including sensor sleep scheduling [43] and power aware cognitive communication protocols [1]. Specifically, in healthcare applications, one of the most important considerations is designing energy efficient protocols.

For instance, interconnecting bio-sensors mounted or embedded on body via Wireless Body Area Network (WBAN) can be implemented with energy efficient Multiple Access Control (MAC) protocol. MAC layer is responsible to manage data packet transmissions from the sensors across the network. Characteristics like scheduling duration of sensor sleep time, path routing and scheduling duty cycles are MAC protocol adjustment methods that can extend battery life of wearables [10].

Chang et. al proposed a routing protocol for WBAN considering expected transmission count and residual energy metrics for optimal path selection [13]. Pradhan et. al compared energy consumption in four protocols, 802.15.4, IEEE 802.15.6, SMAC and TMAC as hybrid MAC protocols in healthcare [34].

Scheduling sleep duration of the sensors are proposed to enhance the energy efficiency by reducing the unnecessary idle listening. Scheduling sleeping periods to match the needs of different applications in sensor networks are proposed[15, 44, 42, 46]. In 2017, Kaur et al. [27] proposed a solution to determine sleep intervals of sensors based on their remaining battery level, usage history, and quality of the measured signal. Zhang et al. [45] reduced the computational complexity by formulating the sleep scheduling while also paying attention to the network reliability in WBANs [45].

Typically, these methods are utilized to improve energy efficiency in a system level, turning attention away from the context. For applications in the healthcare domain, sensory data type, biosignals, sent to the gateway magnifies the importance of data accuracy impacted by both the context of environment and the dynamics of the system itself. Received data in the gateway can be managed using fog-assisted energy efficient optimization algorithms to find optimal solutions for the sensor's configurations to maximize the sensor's battery life while maintaining reliability.

Studies conducted by Zois et al. [47, 48, 49] focused on detecting activity of an individual by gathering partial observations from the sensors considering gateway as the energy bottleneck. However, the objective of these frameworks, based on Markov Decision Process theory, is that of detecting the activity itself, and the focus is on the optimization of data transfer to a sink. In contrast, we propose using the activity as a context to control sensing accuracy in an edge-based architecture to extend the lifetime of sensors.

## 4 Design Challenges

As we discussed earlier, a remote patient monitoring system is expected to make the patients' vital signs available and collectible for healthcare professionals wirelessly. Furthermore, it was mentioned that the PPG signal is a good source for at least three vital signs.

Even though the use of PPG signal is a good source of data, it imposes the system to several challenges. The first challenge is the behavior of the PPG signal in different contexts. The typical PPG signal displayed in Figure 3 is recordable in the hospital setting when the patient is sitting or lying on the bed without movement. This typical PPG signal consists of two parts: AC part which oscillates with each heartbeat and DC part which forms the baseline of the signal. The AC part is the result of the changes in the amount of oxygenated hemoglobin in the blood and the changes in the DC part are due to the pressure applied from other body tissues to the blood vessels. For example, during the respiration process, the change in the size of lungs applies pressure to blood vessels and causes a low-frequency oscillation in the DC part. These kind of changes in the DC part enable us to measure the respiration rate from the DC part oscillation frequency by filtering the AC part. Such changes in the DC part creates a challenge in signal processing when the patient has some activities. Each body movement causes a change in the DC part and more intense activities make larger changes in the signal baseline. When the amplitude of changes is larger than the amplitude of AC part, the detection of heartbeat peaks would be more difficult. In addition, when the body movements are rhythmic (e.g. walking, jogging, running) with a frequency close to respiration, the calculation of respiration rate is also rather difficult.

The other challenge in the PPG signal acquisition is the amount of noise in the signal. Ambient light diffuses to the exposed body tissues close to the sensor spot and causes a level of noise to the recorded signal. Although increasing the brightness of LEDs in the PPG sensor reduces the effect of ambient light noise, it increases also the power consumption in the sensor node.

The last PPG-related challenge is that the most parts of signal processing are not possible to carry out with low power microcontrollers of wearable sensor nodes. The sensor node should send the raw signal to a gateway or cloud server for further processing. This, in turn, requires more power for radio transmission. In the following sections, we describe the solutions to cover the mentioned challenges.

## 5 IoT System Architecture

Internet of Things is a term for describing methods that enable us to sense and control a variety of parameters and objects wirelessly through the Internet. De-veloping a system based on IoT has several benefits in comparison to traditional wireless sensors and remote control approaches. The most significant characteristic of IoT is its well-organized distribution of energy sources, data storage, and processing power. As shown in Figure 4 a common architecture of IoT-based remote health monitoring application consists of three layers. The first layer belongs to the sensor nodes that are recording and reporting the patient state including medical parameters, activity level, posture, location, and environment properties. Sensor nodes send the collected data to one or more gateway devices which are placed close to the patients so that energy cost of radio communication between sensors and gateways remains at a low level. In the gateway layer, the devices have their own storage and processing units powerful-enough to perform pre-processing and fog-computing actions before transferring data to the cloud server. The pre-processing actions may include data filtering, data fusion, data analysis, compression, and encryption. Fog-computing methods may be used to offload a portion of cloud-based tasks to the gateway device at the edge of the network to reduce the bandwidth need, data size, and server load. In the cloud layer, the server receives collected data from several gateways and stores the data recorded from all patients. Such a huge amount of data enables the cloud server to compare patients with their earlier conditions, with other patients in the same condition, and the consequences of the current condition in other patients. The server then would be able to learn from the patients' history and predict the future of patients' health.

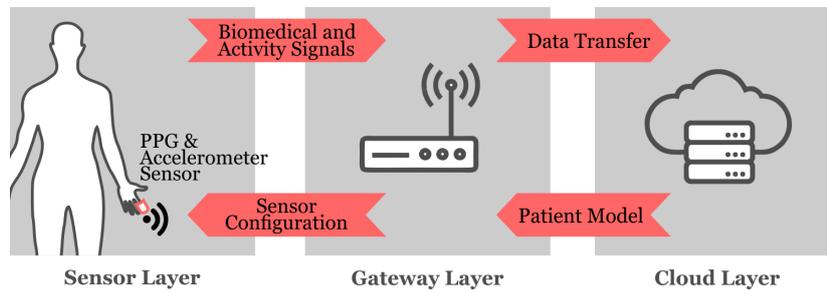

Fig. 4: IoT system architecture

In our setup, we use a battery-powered microcontroller connected to a digital I2C PPG sensor, a 3D accelerometer sensor, a temperature sensor, and a WiFi transmission module as our sensor devices. The amount of power required to drive LEDs in the PPG sensor is configurable, the sensors node is able to receive configurations from the gateway device remotely. The other configurable parameters are the recording and hibernation durations. The gateway is a WiFi-enabled Linux machine which receives the recorded data, performs fog-assisted optimization algorithms, decides about the configuration of the next recording period, and sends the new settings back to the sensor device. The cloud server gets updated with all collected data and updates the RMSE values for the gateway over longer periods of time.

## 5.1 Fog Computing and Its Benefits

Fog computing leverages the concept of Geo-distribution of networks at the edge, enabling local/hierarchical data analysis, decision making, and storage [11, 37, 36]. Prevalence of connected devices and IoT-based systems demands an intermediary layer of computation, in which the local solutions provide low-latency responses, load balancing, and adaptivity for system behavior.

With the growth of IoT-based systems, a rapid increase in the number of connected devices has led to a massive volume of data that needs to be processed [16, 17]. Cloud computing has, thus far, provided scalable and on-demand storage and processing resources to fulfill the requirement of IoT. However, most recent IoT-based applications require mobility, low-latency response, and location-awareness [38]. Moreover, the latency of data transmission between the edge and the cloud is unsatisfactory especially in latency-sensitive systems such as health monitoring [17]. In this regards, Cisco states that "*Today's cloud models are not designed for the volume, variety, and velocity of data that the IoT generates.*" [40]. Therefore, fog computing can be considered as a complementary solution for the cloud computing paradigm to enable such latency demanding applications [31] as it can relocate location dependent, time-dependent, massive scale, and latency-sensitive tasks from the cloud server to the edge of the network [23].

Fog computing provides several lightweight services at the edge of the network, locally analyzing data collected from heterogeneous connected devices. Depending on the computational capacity of the edge servers or gateway devices, such fog-based services can include not only conventional tasks such as protocol conversion but also local data processing applications, some of which are outlined as follows. There is a variety of applications such as data filtering and data fusion to ensure high-level data quality at the edge, improving the data accuracy and performing data abstraction [36, 9]. Such applications can decrease the amount of data that should be sent to the cloud server and subsequently save external bandwidth. Moreover, local decision-making is a solution at the edge by which the system's availability and reliability are increased particularly when the Internet connection is poor [6]. Adaptive sensing and actuation is another application that intelligently tunes the system's configuration at the edge according to the context information [4]. Such a dynamic reconfiguration can considerably improve the system-driven quality attributes such as energy efficiency. Certain security related services can be also performed at the edge, protecting data from unauthorized access (e.g., authentication, data encryption/decryption, anomaly detection, etc.) [36, 39].

In summary, fog computing provides several benefits to the system. The benefits include (but not limited to) i) local processing, notification, and actuation with short latency, ii) interoperability and reconfigurability for the system, iii) energy efficiency for sensor nodes, and, iv) mobility support for the users, and v) reliability and availability of the service. Storing data close to the sensor network and processing it locally leads to quick notification and rapid response. Acting as a power manager, fog computing can reduce and optimize the energy consumption of the sensor nodes while not imposing the management overheads to the sensors.

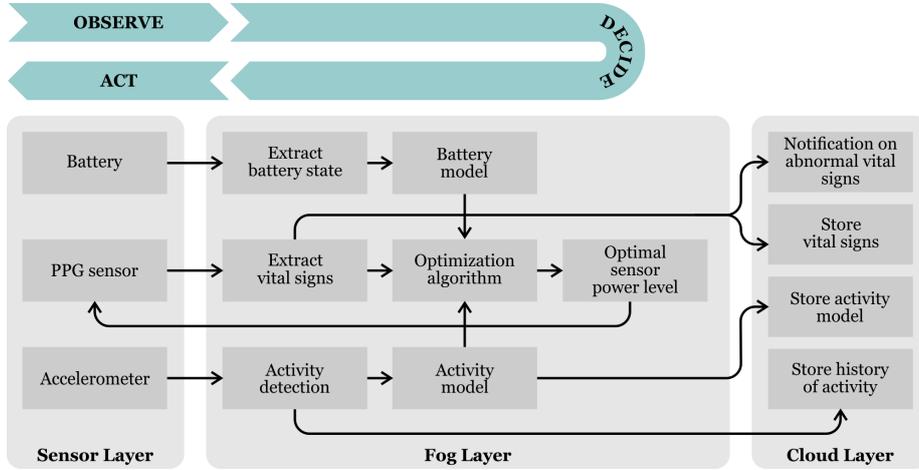

Fig. 5: The high level system architecture

## 6 Fog-assisted Runtime Energy Management in Wearable Sensors

Based on the design challenges discussed in Section 4, it is desired to create a platform that leverages portable sensors coupled with a layered communication and processing architecture for local control purposes. Accommodating runtime optimization algorithms at the edge enables the deployment of smart solutions to control the settings of the sensors. Configurations/knobs such as sampling frequency and transmission rate can also be used to control the communication delay and bandwidth between sensor and fog layer. However, in this chapter, we only focus on controlling the sensor's sensing power to minimize energy which does not influence the communication characteristics of the system (i.e., the volume of transmitted data is unchanged). Energy efficient algorithms need to consider the environmental situation to find a robust and optimal solution. Daily activities, in particular strenuous ones, impact the quality of sensor's measurements. For instance, the noise caused by body movements during running is strong enough to significantly distort the PPG signal. In contrast, physical activities such as, sleeping or sitting impose less motion artifacts. Therefore, smaller energy budgets can be assigned to the sensor to satisfy similar threshold of accuracy. This highlights the necessity of context-awareness in managing quality and energy of characteristics of sensors at the edge.

The high level view of our proposed fog-assisted architecture is shown in Figure 5. The architecture consists of three different phases as Observe, Decide and Act.

- Observe: Collecting data from a PPG sensor and an Accelerometer along with the battery state of charge of the sensor node and transmitting it wirelessly to the fog layer.

- Decide: i) Preprocessing and filtering the data received from the sensor layer, ii) extracting vital signs from the PPG signal, iii) detecting activity of a user from collected Accelerometer data, iv) modeling the statistics of activity dynamics, v) calculating the transition probabilities between activity states, vi) modeling transition probabilities between battery states during charging and discharging modes, and vii) implementing optimization algorithms to utilize the user's activity model, battery transitioning model, and vital signs to find an optimal solution for the sensor's power level.

- Act: Regulating the sensor node's power level based on the optimal solution derived in the fog layer.

In addition, the activity model, vital signs, and the user's history of activity are stored in the cloud for further analysis. Notifications on detecting abnormalities in vital signs are transmitted to the cloud to warn the clinicians. Section 6.1 discusses the concept of computational self-awareness in the proposed IoT architecture, using the three processes of Observe, Decide and Act, while Section 6.2 presents the proposed energy optimization algorithms to control the sensor layer.

## 6.1 Computational Self-awareness

Computational self-awareness enables a computing system to act reliably, optimally, and adaptively, despite the radical changes in its own state and environment [29, 28, 35, 18]. Computational self-awareness therefore requires the computing system to be empowered by knowledge about both itself as well as the surrounding environment. By exploiting computational self-awareness principles, the system's dynamic behavior can be managed to provide a high-level of quality of service, allowing accurate optimization schemes. Computational self-awareness has thus far been investigated in various applications including cyber-physical systems, remote health monitoring and, mobile applications [29, 5, 7].

Within healthcare applications, the system leverages semantic information including the synergy between the system, the individual, and the surrounding environment. In this regard, self-awareness is performed in computing systems via a closed-loop framework where three different phases – Observe, Decide, and Act (ODA) – are deployed together with reflective models of the system [22, 19]. (Figure 5) shows the application of the ODA loop, together with reflective models of the battery and activity, in the IoT-based health monitoring system architecture. Using the self-aware enhanced ODA-loop, the cognitive fog layer adaptively configures the sensor nodes to provide a high-level of accuracy as well as enhancing the energy and bandwidth efficiency of the system.

In each iteration, first, data collection is performed via the sensor network, obtaining internal (e.g., system status) and external (e.g., user's health status) data. Second, the collected data are fed to data analytic approaches and models are created. Then, the best behavior (i.e., the system's configuration) is determined according to the recent observation. The selected configuration should satisfy both the system's and application's requirements. Third, if adaptation in system behavior is needed, the changes are applied, and the selected configuration is implemented.

## 6.2 Energy Optimization Algorithms

In our proposed IoT system architecture, the fog layer is equipped with a cognitive optimization algorithm that tracks the dynamics of the sensor and user's activity to offer an optimal solution for the sensor's energy level. Our optimization formulation centers around the *minimization of energy expenditure of a sensor node while satisfying the application requirements in terms of measurement accuracy.* Measurement accuracy is directly influenced by the power level chosen in the sensor. Figure 6 demonstrates the process of modeling accuracy measurement as a function of power level in a PPG sensor. We first start by introducing the error in the accuracy of measurements. We use *Root Mean Squared Error*($RMSE$) between vital signs extracted from PPG signal (e.g., $SpO_2$, heart rate, and respiration rate) and the true values of the extracted features. The ground-truth features are extracted from the following three sensors:

- A chest band using an ECG sensor to collect the reference heart-rate

- An airflow sensor placed on the user's upper lip to measure the true value of respiratory rate

- A precise PPG sensor with higher power consumption to extract $SpO_2$

We set the PPG sensor's power level (e.g., $U$) to all possible values. Higher power levels lead to more energy consumption as well as higher accuracy of measurements. In order to incorporate context-awareness in our problem formulation, we assume users are engaged in different activities. Based on the intensity of the activity, a different amount of motion artifact will be observed in the signal. We consider the following list of activities: "Sleeping", "Sitting", "Walking", "Jogging" and "Running". $RMSE$ is calculated for different combinations of the sensor's power level for different user's activities (e.g., $X$). Comparing the true values of the vital signs as references to the measured values is a proper metric to model the accuracy. Therefore, the proposed approach can model the error in the accuracy of extracted vital signs (e.g., $\epsilon(X, U)$) as a function of sensing power level in the sensor and the activity of an individual.

Consider that the Probability Density Function (PDF) of error denoted by $\rho(\epsilon(X, U))$ to follow a Gaussian distribution. Therefore, we can define the probability of error as the tail probability of normal distribution when the user's activity $X = x$ and the power level of the sensor is $U = u$ where,

$$\mathcal{P}_\tau(X, U) = \int_{-\infty}^{\tau} \rho(\epsilon(X, U) \mid X = x, U = u) d\epsilon. \qquad (3)$$

The threshold $\tau$ determines the maximum threshold in error tolerance. For instance, larger values of $\tau$ shows that larger values of $RMSE$ is acceptable in calculating the error probability $\mathcal{P}_\tau(X, U)$. We now define the regions of abnormal vital signs by marking the vital signs $y$ as abnormal and assign a Gaussian PDF with calculated mean $m_a$ and variance $\sigma_a$ (e.g., $f_a(y|X) \sim \mathcal{N}(m_a, \sigma_a)$). Herein, the probability of abnormal vital signs ubiquity can be distinguished from the normal vital signs with threshold $\theta$. Considering the user's activity

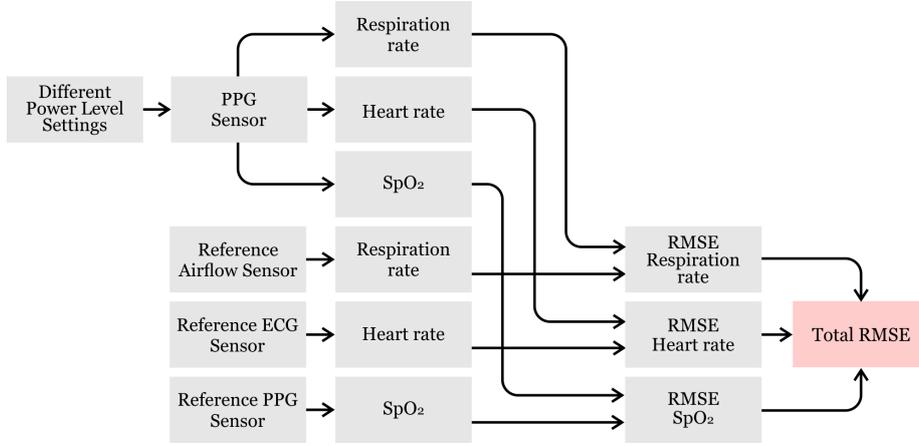

Fig. 6: Modeling accuracy of measurements (e.g. $\epsilon(X, U)$) in PPG sensor using $RMSE$.

$X = x$, the probability of existence of abnormal vital signs can be written as,

$$\mathcal{P}_\theta(X) = \int_\theta^\infty f_a(y \mid X = x) dy. \tag{4}$$

Equation 3 along with the marked abnormal vital signs in Equation 4 can be used to determine the probability of abnormality misdetection. The probability of abnormality misdetection is a metric to determine the possibility that abnormal vital signs are not detected due to error in sensor's measurements. In fact, probability of misdetection can be defined as a joint event of ubiquity of abnormal vital signs and sensor's error tolerance. We proved in [4] that upper bound for probability of misdetection in abnormality (e.g. $\mathcal{P}_D$) can be written as,

$$\mathcal{P}_D(X, U) = \mathcal{P}_\theta(X) \mathcal{P}_\tau(X, U). \tag{5}$$

We use formulation of $\mathcal{P}_D(X, U)$ as the main presentation of deterioration risk factor. After modeling the probability of misdetection, we can define the important factors to be optimized. On one hand, lower power levels in the sensor consumes less energy leading to the ability to monitor a patient for a longer time. However, the signal captured by the sensor is distorted by noise due to low Signal to Noise Ratio (SNR). On the other hand, using higher energy levels in the sensor increases the energy consumption leading to shorter battery life, but enhances the accuracy of the extracted vital signs. Therefore, it is important to find an optimal solution as a trade off between energy consumption (e.g., $\mathcal{C}_{\text{TX}}(U)$) as a result of choosing power level $U$ and desired level of satisfactory in probability of misdetection. The optimization problem can be defined as:

(a) minimizing cost function of energy consumption over constraints of probability of misdetection;

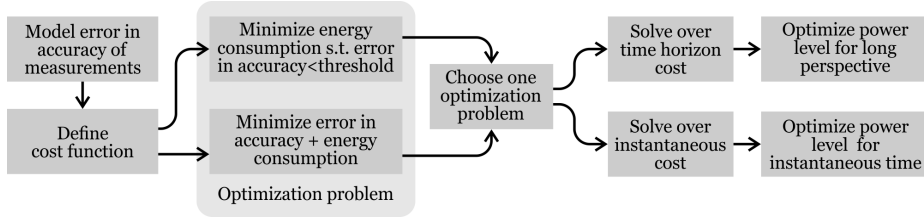

Fig. 7: Optimization algorithm implementation.

$$\begin{aligned}
\underset{U}{\text{minimize}} \quad & \mathcal{C}_{\text{TX}}(U) \\
\text{subject to} \quad & \mathcal{P}_{\text{D}}(X,U) \leq \eta \\
& \text{or, equivalently,} \\
& \mathcal{P}_{\tau}(X,U) \leq \frac{\eta}{\mathcal{P}_{\theta}(X)} = \zeta
\end{aligned} \qquad (6)$$

or *(b)* defining total cost function of energy consumption and probability of misdetection with parameter $0 \leq \omega \leq 1$ to control the importance weight of the two factors;

$$\mathcal{C}_{\text{total}}(X,U) = \omega \mathcal{P}_{\text{D}}(X,U) + (1-\omega)\mathcal{C}_{\text{TX}}(U). \qquad (7)$$

Therefore, the optimization problem can be written as:

$$\underset{U}{\text{minimize}} \quad \mathcal{C}_{\text{total}}(X,U) \qquad (8)$$

Defining an optimization problem based on the cost function requires finding an optimal solution for the power level of the sensor. Solutions to the optimization problems can be proposed to minimize the accumulated cost function over finite time horizon or minimizing the instantaneous cost. Therefore, both methods find an optimal power level for the sensor. Figure 7 demonstrates the process of implementing the optimization algorithm.

Two approaches are proposed in this chapter to solve the optimization problem, (i) *myopic* strategy and (ii) Markov Decision Processes (MDP) strategy. In addition, the performance of aforementioned strategies are compared.

### 6.2.1 Myopic Strategy

Optimizing the instantaneous cost function results in implementing real-time algorithms with linear time complexity known as the *myopic* strategy. Intuitively, myopic strategy finds an optimal solution for the sensor's power level based on instantaneous activity states. Myopic strategy can find the lowest power level in Equation 6 that satisfies the maximum probability of misdetection. This strategy can be implemented on the edge devices with limited memory and computations to calculate real-time solution for the sensor's power level. In order to find a solution based on a longer perspective to avoid possible outage and performance reduction, a strategy that evaluates the outcome of all possible actions can be proposed. Next section discusses MDP strategy that allows optimization of power level over a temporal horizon.

### 6.2.2 MDP Strategy

Optimizing the accumulated cost over time horizon can be achieved by tracking the history of dynamics of the system as well as exploring different possible actions to achieve longer perspective of solutions. MDP is a strategy that calculates the optimal power level in the sensor by using a sliding window over a finite time horizon by tracking the stochastic state of the system. Stochastic model in MDP consists of user's activity as the contextual state and battery's state of charge as the self state. We consider that the battery is quantized as distinct battery states. We model the transitioning probabilities between battery states depending on a possible power level chosen as the action. Finite battery states can be defined as $Q = \{Q_1, Q_2, ..., Q_K\}$ with $Q_1$ as the battery state completely discharged and $Q_K$ as the fully charged state. We can define transitioning probabilities from battery state $Q_k$ to $Q_{k'}$ with the possible discrete power levels $U \in \{U_0, U_1, U_2, ..., U_m\}$,

$$q(k'|k, U) \triangleq \mathcal{P}\left(Q'_k | Q_k, U\right) \tag{9}$$

Considering the transition probabilities defined in Equation 9, the dynamics of the battery can be modeled as a Markovian process with the states transitioning shown in Figure 8.

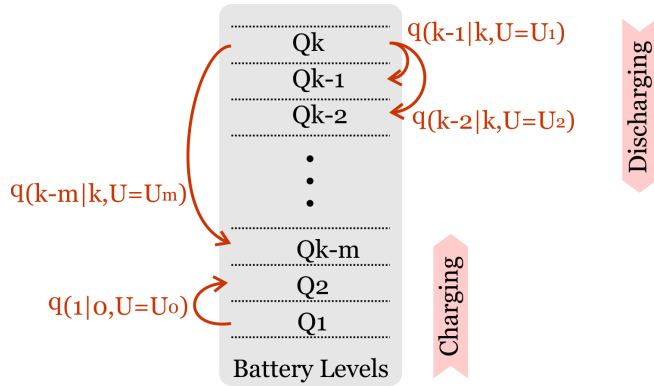

Fig. 8: Markov chain of battery states during charging and discharging.

In addition, Markov model tracks the changes in activity of a user as state space $X$ in {"Sleeping", "Sitting", "Walking", "Jogging", "Running"}. Note that activities during the day change over time and building different models is a necessity. Therefore, we uniformly break down the daily activity into $n$ smaller periods. We consider Markov model corresponding to $i$th period in $\{1, 2, ..., n\}$ each with duration of $\frac{24}{n}$ hours.

Figure 9 shows the example of Markov chain for activities during one period. This type of model is updated to the change of subject's context over time and personalized based on the daily activity. We build the Markov chain transitioning from activity $X_j$ to $X_{j'}$ during period $i$ with the following transition probabilities,

$$p_i(j'|j) = \mathcal{P}\left(X_{ij'} | X_{ij}\right) \tag{10}$$

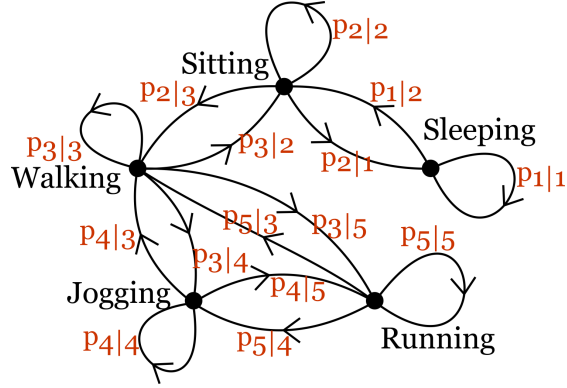

Fig. 9: Markov chain of activities of an individual during one period.

The total state of the system can be written as the joint event state space of battery state $Q_k$ and activity state $X_{ij}$. We prove that the transition probabilities in joint states is the multiplication of battery and activity transition probabilities [3]. The resulting Markov chain with proposed state transitions considering the changes of activity as a function of the time of the day is shown in Figure 10.

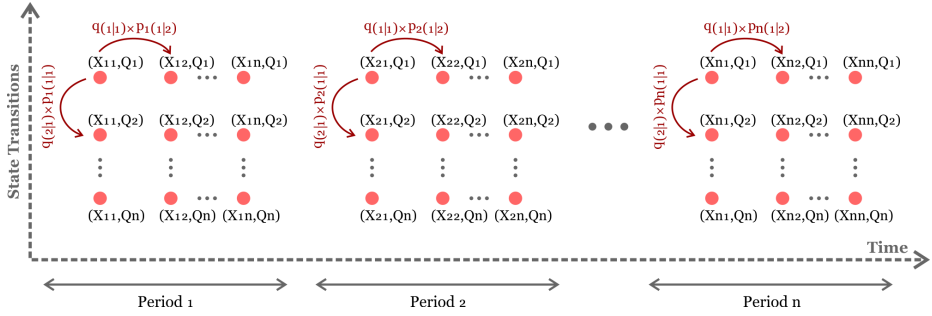

Fig. 10: Markov chain of joint battery and activity states during period 1 to $n$.

We take advantage of the non-homogeneous Markovian model, where transition probabilities are calculated depending on the period of the day. The optimization algorithm then chooses an optimal action $U^*$ such that the accumulated cost function over the finite time horizon $N$ can be calculated based on the instantaneous cost function. Herein, with problem formulation in Equation 7 we have:

$$U^* = \underset{U}{\mathrm{argmin}}\, \mathbb{E}\left(\sum_{t=0}^{N} \gamma^t \left[\mathcal{C}_{\text{total}}\right]\right) \quad (11)$$

Where, $\gamma$ is the discount factor taking values commonly between 0.9 and 1.

Figure 11 shows the comparison between two strategies *myopic* with problem formulation presented in Equation 6 and MDP with problem formulation shown in equation 8 for a subject during 24 hours of monitoring. The daily activity of the subject is divided into 4 periods of 6 hours (e.g., $i = \{1, 2, 3, 4\}$). For each period, the transition probability between activity is trained based on the history gathered from one week of activity monitoring. Markov chain for each period is modeled based on the transition probabilities shown in Figure 9. Since the activity model of a user changes over time, the transition probabilities need to be updated regularly. We calculated the average transition probabilities for each period over one week of activity data. The trained model is used for the following week to find the optimal solution. In addition, we evaluated our results for activities of 14 subjects. For each subject, the model is updated weekly to adapt the system to its change throughout the time. The results are evaluated for four weeks of data to calculate average energy consumption in the PPG sensor layer using *myopic* and MDP strategies. The joint battery and activity states are modeled and the corresponding Markovian model based on the period of day is used to find an optimal solution in MDP strategy.

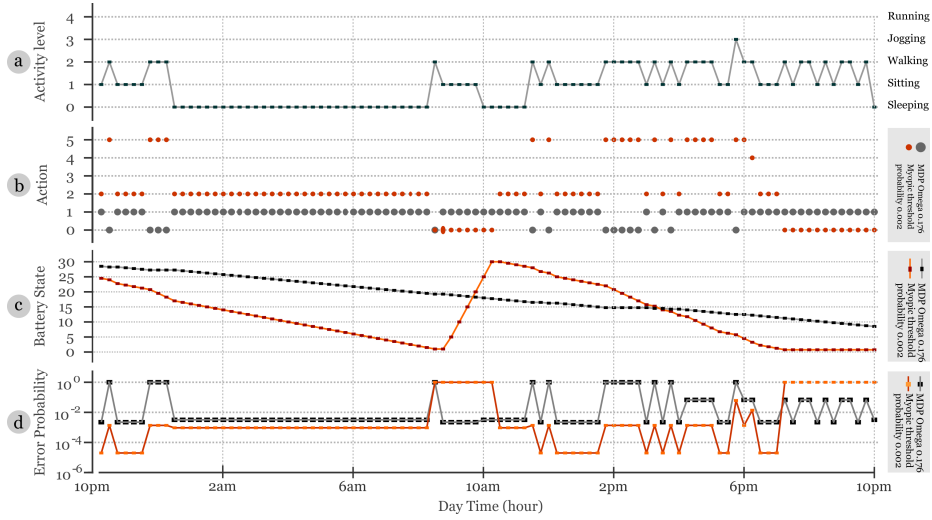

Fig. 11: 24-hour health monitoring of a healthy person . (a) User's activity level. (b) Optimal sensor's power level. (c) battery state tracking based on sensor's power level. (d) Probability of error expected regarding the user's activity. The red line, and grey line indicate the *myopic* and MDP methods, respectively.

We compared MDP and *myopic* strategies by setting the parameter $\omega$ in Equation 7 to 0.17 and the threshold for probability of error to 0.002 (e.g., $\zeta$ in Equation 6). For the sake of fair comparison, the parameters $\omega$ and $\zeta$ are chosen in the way that average probability of misdetection over four weeks for both *myopic* and MDP methods become the same. Parameter $\omega$ determines the importance of optimizing total cost function in MDP. For values close to 0, the energy consumption will have higher impact on the cost function. Therefore, MDP chooses policies to minimize energy consumption in the sensor. Whereas, $\omega$ close to 1 stress the importance of probability of misdetection in Equation 7. In this case, MDP chooses the optimal actions to fulfill minimization of probability of misdetection. During one month of monitoring, the fixed average probability of error will achieve 0.26 by setting $\omega = 0.17$ in MDP and $\zeta = 0.002$ in *myopic* strategies. Figure 11 (a) shows the activity of this subject for 24 hours starting at 10 pm to 10 PM of the following day. Figure 11 (b) shows comparison of optimal actions taken using the two proposed methods. This sensor's sensing power level could be chosen in 5 different settings, $U \in \{U_0, U_1, U_2, U_3, U_4, U_5\}$. Power levels in aforementioned PPG sensor can be used with corresponding current levels (e.g., $U_1 = 0.8$ mA, $U_2 = 3.5$ mA, $U_3 = 6.3$ mA, $U_4 = 9.2$ mA and $U_5 = 12$ mA). For each current level, the power consumption is specified. For instance, current level $U_0$ indicates the sensor is in the sleeping mode. The power consumption is {69.3 mW, 73.26 mW, 79.86 mW, 84.15 mW, 89.43 mW}, respectively for the rest of the current levels. We set the battery states so that the PPG sensor reduces 1 level of state during 1 hour of sensing when the current level $U_1$ is chosen. The highest current level $U_5$ consumes 5 levels of battery during one hour. Therefore, the battery is discharged after 30 hours of using the lowest current level $U_1$. Figure 11 (c) shows the battery state tracking using the optimal actions in both strategies with $Q = \{Q_1, Q_2, ..., Q_{30}\}$ as the battery states. The optimal action changes the battery state over time. It is shown that MDP strategy can extend the battery lasting twice more compared to *myopic* method. This is useful since the frequency of charging the battery is less resulting in planning wisely on using the energy resources. Figure 11 (d) demonstrates the probability of error with respect to the action taken in both methods. Note that it is assumed that the sensor will be forced to go to sleeping mode $U_0$ when the battery is discharged which results in probability of error to be 1. *Myopic* method finds an optimal solution just based on the current activity of a user without considering that the chosen action will discharge the battery faster and this will increase the likelihood the battery drainage during vigorous activities including, walking, jogging, and running which requires accurate monitoring. *Myopic* has a linear time complexity leading to solving the optimization problem in real-time. This method specifically is practical when the model of activity of a subject is not available for training Markovian model. During one month of monitoring, we observed the average of 12% reduction in energy consumption comparing MDP with *myopic* fulfilling the same probability of error. The increase in battery lasting was observed with 2× in MDP.

Figure 12 compares two methods of MDP, *myopic* and selection of constant power level. In this experiment, average error probability is calculated as a function of average energy consumption during one month for 14 subjects. Results indicate that in these three methods, choosing parameters in MDP strategy can save energy compared to *myopic* method. For instance, for a fixed average error probability of 0.32, MDP consumes 3.8 KJ, while *myopic* consumes 4.46 KJ and static power level with $U_4 = 9.2$ mA. Therefore, MDP can observe average of 12% reduction in energy consumption while fulfilling the same error probability. In addition, highest power level $U_5 = 12$ mA consumes slightly less power while the average error probability of 0.25 is achieved. Our model takes into consideration that the user charges the battery with probability of one only when they are sleeping or sitting but not walking, jogging or running. Choosing high power levels leads to a faster drainage of battery meaning that the user needs to recharge the battery more frequently. During the charging process, the probability of error is 1. Therefore, the average probability of error in $U_5 = 12$ mA is higher than using $U_4 = 9.6$ mA during the one month of monitoring the 14 subjects.

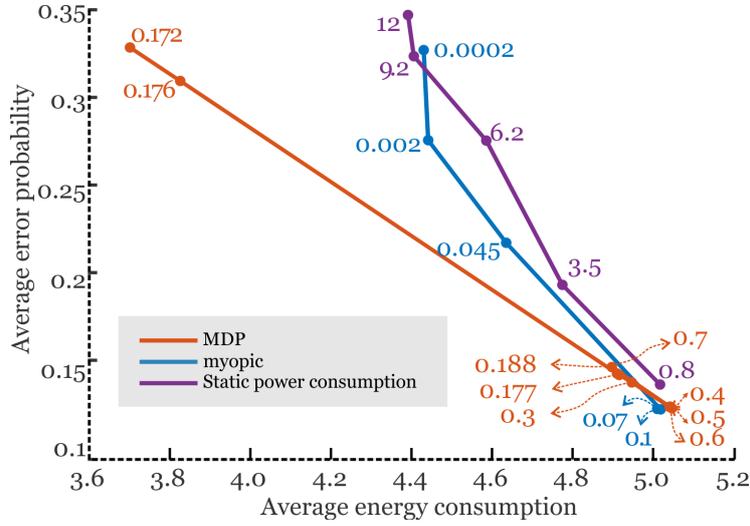

Fig. 12: Average probability of error as a function of energy consumption (KJ) in four weeks averaged over 14 subjects. Comparison between three methods of MDP, *myopic* and static power consumption. Myopic method is evaluated based on $\zeta \in \{0.0002, 0.002, 0.045, 0.07, 0.1\}$ in Eq. 6. MDP is evaluated based on $\omega \in \{0.172, 0.176, 0.177, 0.188, 0.3, 0.4, 0.5, 0.6, 0.7\}$ in Eq. 8. Static power consumption is calculated based on $U \in \{0.8 \text{ mA}, 3.5 \text{ mA}, 6.2 \text{ mA}, 9.2 \text{ mA}, 12 \text{ mA}\}$

## 7 Conclusions

The 3-layer IoT paradigm has opened new avenues of monitoring patients out of hospital using wearable sensors. Such a system includes wearables with tight energy constrains making it critical to empower these sensors with runtime energy management approaches. User context can be exploited to minimize the measurement energy with a minimal loss of measurement accuracy, however, such algorithms require contextual information to control the energy budget in the sensor to monitor subjects accurately imposing energy overhead on the sensor layer. Energy-constrained wearable sensors cannot often afford such an overhead, however, if a proper architecture is used, the overhead can be migrated to the next layer (i.e., smart gateways at the fog layer) enabling local fog-assisted control of sensors. The fog layer provides an opportunity to perform real-time control of the sensor's configurations. In this chapter, we demonstrated optimization methods to address a two fold goal, *minimizing the energy consumption while fulfilling satisfactory threshold of probability of misdetection in abnormality*. We used the key idea of context-awareness to bring accurate solutions to the optimization algorithms. We proposed two methods, optimizing instantaneous cost function known as, *myopic* strategy and MDP as the solution to accumulated cost function over finite time horizon. We compared results of MDP and *myopic* during monitoring a subject for 24 hour and we observed average of $2x$ battery life extension in MDP strategy compared with *myopic* strategy.

## Acknowledgement

This material is based upon work supported partially by the US National Science Foundation (NSF) WiFiUS grant CNS-1702950 and Academy of Finland grants 311764 and 311765.

## References

[1] Adnan Aijaz and A Hamid Aghvami. "Cognitive machine-to-machine communications for Internet-of-Things: A protocol stack perspective". In: *IEEE Internet of Things Journal* 2.2 (2015), pp. 103–112.

[2] John Allen. "Photoplethysmography and its application in clinical physiological measurement". In: *Physiological measurement* 28.3 (2007), R1.

[3] Delaram Amiri et al. "Context-Aware Sensing via Dynamic Programming for Edge-Assisted Wearable Systems". In: *ACM Transactions on Computing for Healthcare*. 2019.

[4] Delaram Amiri et al. "Edge-Assisted Sensor Control in Healthcare IoT". In: *2018 IEEE Global Communications Conference: Selected Areas in Communications: E-Health (Globecom2018 SAC EH)*. Abu Dhabi, United Arab Emirates.

[5] Arman Anzanpour et al. "Self-awareness in remote health monitoring systems using wearable electronics". In: *Proceedings of the Conference on Design, Automation & Test in Europe*. European Design and Automation Association. 2017, pp. 1056–1061.

[6] Iman Azimi et al. "Hich: Hierarchical fog-assisted computing architecture for healthcare iot". In: *ACM Transactions on Embedded Computing Systems (TECS)* 16.5s (2017), p. 174.

[7] Iman Azimi et al. "Self-aware early warning score system for IoT-based personalized healthcare". In: *eHealth 360*. Springer, 2017, pp. 49–55.

[8] Mirza Mansoor Baig and Hamid Gholamhosseini. "Smart health monitoring systems: an overview of design and modeling". In: *Journal of medical systems* 37.2 (2013), p. 9898.

[9] Paolo Bellavista et al. "A survey on fog computing for the Internet of Things". In: *Pervasive and Mobile Computing* (2018).

[10] Chintan Bhatt, Nilanjan Dey, and Amira S Ashour. "Internet of things and big data technologies for next generation healthcare". In: Springer, 2017.

[11] Flavio Bonomi et al. "Fog computing and its role in the internet of things". In: *Proceedings of the first edition of the MCC workshop on Mobile cloud computing*. 2012.

[12] Luca Catarinucci et al. "An IoT-aware architecture for smart healthcare systems". In: *IEEE Internet of Things Journal* 2.6 (2015), pp. 515–526.

[13] Lin-Huang Chang et al. "Energy-efficient oriented routing algorithm in wireless sensor networks". In: *Systems, Man, and Cybernetics (SMC), 2013 IEEE International Conference on*. IEEE. 2013, pp. 3813–3818.

[14] Peter H Charlton et al. "An assessment of algorithms to estimate respiratory rate from the electrocardiogram and photoplethysmogram". In: *Physiological measurement* 37.4 (2016), p. 610.

[15] Yunxia Chen et al. "Transmission scheduling for optimizing sensor network lifetime: A stochastic shortest path approach". In: *IEEE Transactions on Signal Processing* 55.5 (2007).

[16] Rudyar Cortésa et al. "Stream Processing of Healthcare Sensor Data: Studying User Traces to Identify Challenges from a Big Data Perspective". In: *Procedia Computer Science* 52 (2015), pp. 1004–1009.

[17] Amir Vahid Dastjerdi and Rajkumar Buyya. "Fog Computing: Helping the Internet of Things Realize Its Potential". In: *Computer* 49.8 (2016), pp. 112–116.

[18] Anind K Dey, Gregory D Abowd, and Daniel Salber. "A conceptual framework and a toolkit for supporting the rapid prototyping of context-aware applications". In: *Human–Computer Interaction* 16.2-4 (2001), pp. 97–166.

[19] Nikil Dutt, Axel Jantsch, and Santanu Sarma. "Toward smart embedded systems: A self-aware system-on-chip (soc) perspective". In: *ACM Transactions on Embedded Computing Systems (TECS)* 15.2 (2016), p. 22.

[20] Ainara Garde et al. "Estimating respiratory and heart rates from the correntropy spectral density of the photoplethysmogram". In: *PloS one* 9.1 (2014), e86427.

[21] Jayavardhana Gubbi et al. "Internet of Things (IoT): A vision, architectural elements, and future directions". In: *Future generation computer systems* 29.7 (2013), pp. 1645–1660.

[22] Henry Hoffmann et al. *Seec: A framework for self-aware computing*. Tech. rep. 2010.

[23] Hua-Jun Hong. "From Cloud Computing to Fog Computing: Unleash the Power of Edge and End Devices". In: *2017 IEEE International Conference on Cloud Computing Technology and Science (CloudCom)*. 2017, pp. 331–334.

[24] "Internet-of-Things and big data for smarter healthcare: From device to architecture, applications and analytics". In: *Future Generation Computer Systems* 78 (2018), pp. 583 –586.

[25] SM Riazul Islam et al. "The internet of things for health care: a comprehensive survey". In: *IEEE Access* 3 (2015), pp. 678–708.

[26] Walter Karlen et al. "Multiparameter respiratory rate estimation from the photoplethysmogram". In: *IEEE Transactions on Biomedical Engineering* 60.7 (2013), pp. 1946–1953.

[27] Navroop Kaur and Sandeep K Sood. "An energy-efficient architecture for the Internet of Things (IoT)". In: *IEEE Systems Journal* 11.2 (2017), pp. 796–805.

[28] Peter R Lewis et al. "A survey of self-awareness and its application in computing systems". In: *2011 Fifth IEEE Conference on Self-Adaptive and Self-Organizing Systems Workshops*. IEEE. 2011, pp. 102–107.

[29] Peter R Lewis et al. *Self-Aware Computing Systems*. Springer, 2016.

[30] L-G Lindberg, H Ugnell, and PÅ Öberg. "Monitoring of respiratory and heart rates using a fibre-optic sensor". In: *Medical and Biological Engineering and Computing* 30.5 (1992), pp. 533–537.

[31] Yiming Liu et al. "Distributed Resource Allocation and Computation Offloading in Fog and Cloud Networks With Non-Orthogonal Multiple Access". In: *IEEE Transactions on Vehicular Technology* 67.12 (2018), pp. 12137–12151.

[32] Maxim Integrated. https://www.maximintegrated.com/en/products/sensors/MAX30102.html. (accessed 2018-11-01).

[33] Marco AF Pimentel, Peter H Charlton, and David A Clifton. "Probabilistic estimation of respiratory rate from wearable sensors". In: *Wearable electronics sensors*. Springer, 2015, pp. 241–262.

[34] Gitanjali Pradhan, Rajni Gupta, and Suparna Biswasz. "Study and simulation of WBAN MAC protocols for emergency data traffic in healthcare". In: *2018 Fifth International Conference on Emerging Applications of Information Technology (EAIT)*. IEEE. 2018, pp. 1–4.

[35] Jürgo S Preden et al. "The benefits of self-awareness and attention in fog and mist computing". In: *Computer* 48.7 (2015), pp. 37–45.

[36] Amir M Rahmani et al. "Exploiting smart e-Health gateways at the edge of healthcare Internet-of-Things: A fog computing approach". In: *Future Generation Computer Systems* 78 (2018), pp. 641–658.

[37] Amir M Rahmani et al. *Fog computing in the internet of things: Intelligence at the edge*. Springer, 2017.

[38] Subhadeep Sarkar and Sudip Misra. "Theoretical modelling of fog computing: a green computing paradigm to support IoT applications". In: *IET Networks* 5.2 (2016), pp. 23–29.

[39] Syed Noorulhassan Shirazi et al. "The extended cloud: Review and analysis of mobile edge computing and fog from a security and resilience perspective". In: *IEEE Journal on Selected Areas in Communications* 35.11 (2017), pp. 2586–2595.

[40] Cisco Systems. "Fog Computing and the Internet of Things: Extend the Cloud to Where the Things Are". In: *Available online: www.cisco.com (accessed on 21 March 2019)* (2015).

[41] "The Internet of Things for basic nursing care—A scoping review". In: *International Journal of Nursing Studies* 69 (2017), pp. 78 –90.

[42] Kun Wang et al. "Green industrial internet of things architecture: An energy-efficient perspective". In: *IEEE Communications Magazine* 54.12 (2016), pp. 48–54.

[43] Lan Wang and Yang Xiao. "A survey of energy-efficient scheduling mechanisms in sensor networks". In: *Mobile Networks and Applications* (2006).

[44] Jason L Williams et al. "Approximate dynamic programming for communication-constrained sensor network management". In: *IEEE Transactions on signal Processing* (2007).

[45] Rongrong Zhang et al. "Energy-Efficient Sleep Scheduling in WBANs: from the Perspective of Minimum Dominating Set". In: *IEEE Internet of Things Journal* (2018).

[46] Chunsheng Zhu et al. "Collaborative location-based sleep scheduling for wireless sensor networks integratedwith mobile cloud computing". In: *IEEE Transactions on Computers* 64.7 (2015), pp. 1844–1856.

[47] Daphney-Stavroula Zois, Marco Levorato, and Urbashi Mitra. "A POMDP framework for heterogeneous sensor selection in wireless body area networks". In: *2012 Proceedings IEEE INFOCOM*. IEEE. 2012, pp. 2611–2615.

[48] Daphney-Stavroula Zois, Marco Levorato, and Urbashi Mitra. "Active classification for pomdps: A kalman-like state estimator". In: *IEEE Transactions on Signal Processing* 62.23 (2014), pp. 6209–6224.

[49] Daphney-Stavroula Zois, Marco Levorato, and Urbashi Mitra. "Energy-efficient, heterogeneous sensor selection for physical activity detection in wireless body area networks". In: *IEEE Transactions on signal processing* 61.7 (2013), pp. 1581–1594.